\documentclass[12pt]{article}
\usepackage{amssymb}
\usepackage{axodraw}

\newcommand\be{\begin{equation}}
\newcommand\ee{\end{equation}}

\newcommand\no{\nonumber\\}
\newcommand\bea{\begin{eqnarray}}
\newcommand\eea{\end{eqnarray}}
\newcommand\tet{\theta}
\newcommand\la{\lambda}
\newcommand\lb{\bar\lambda}
\newcommand\m{\mu}
\newcommand\n{\nu}
\newcommand\al{\alpha}
\newcommand\bet{\beta}
\newcommand\ga{\gamma}
\newcommand\de{\delta}
\newcommand\s{\sigma}
\newcommand\e{\epsilon}

\newcommand\ro{\rho}
\newcommand\ka{\kappa}

\newcommand\fr{\frac}

\newcommand\ps{\slash\!\!\!\partial}
\newcommand\del{\partial}
\newcommand\h{\hat}
\newcommand\sll{\slash\!\! l}
\newcommand\slp{\slash\!\!\! p}
\newcommand\slk{\slash\!\!\! k}
\newcommand\sla{\slash\!\!\!\! A}
\newcommand\sld{\slash\!\!\!\! D}

\newcommand\tmn{\theta^{\mu\nu}}
\newcommand\trs{\theta^{\rho\sigma}}

\newcommand{\ssc}{\stackrel{C}{\star}}
\newcommand{\sst}{\stackrel{\theta}{\star}}
\newcommand\dss{\delta_S}
\newcommand\ti{\tilde}

\begin{document}

\vspace{13mm}
\begin{center}
{\Huge \bf N=1/2 Supersymmetric gauge theory in noncommutative
space}

\vspace{17mm}

\"{O}mer F. Dayi$^{a,b,}$\footnote{E-mail addresses: dayi@itu.edu.tr and
dayi@gursey.gov.tr.} \,{and}\, 
Lara T. Kelleyane$^{a,}$\footnote{E-mail address: kelleyane@itu.edu.tr.}\\
\vspace{5mm}

\end{center}

\noindent {\em $^{a}${\it Physics Department, Faculty of Science and
Letters, Istanbul Technical University, TR-34469 Maslak--Istanbul, Turkey.} }

\vspace{3mm}

\noindent {\em $^{b}${\it Feza G\"{u}rsey Institute, P.O. Box 6,
TR--34684, \c{C}engelk\"{o}y--Istanbul, Turkey. } }

\vspace{2cm}

{\small
 A formulation of (non--anticommutative) N=1/2 supersymmetric $U(N)$ gauge theory in noncommutative space is studied. We show that at one loop
  UV/IR mixing occurs. A generalization of
Seiberg--Witten map to noncommutative and  non--anticommutative superspace  is
employed to obtain an action in terms
of commuting fields at  first order in the noncommutativity
parameter $\tet .$ This leads to  abelian and non--abelian gauge theories whose
supersymmetry transformations are
local and non--local, respectively.}

\newpage

\section{Introduction}

Deformation of superspace where fermionic coordinates are non-anticommuting 
appeared in some
different contexts 
\cite{ov}--\cite{bie}. 
At the start 
one can simultaneously deform bosonic
coordinates allowing them to be noncommuting, in terms of a star product embracing both
of the deformations\cite{ty}. 
However, as far as gauge theories are concerned usually
non-anticommutativity is considered alone.
Instead of introducing noncommutativity of bosonic coordinates and
non--anticommutativity of fermionic ones simultaneously from the beginning, 
we may  do it in two steps:  N=1/2
supersymmetric gauge theory action in components
includes ordinary fields and
non--anticommutativity parameter\cite{sei}.  Thus its noncommutative generalization can be
obtained as usual. 
However, the same action would result using the superfield formulation
given in \cite{ty}. Hence, two approaches are
equivalent.
We study this non--anticommutative as well as
noncommutative  theory.

One of the most important features arising in
field theories in noncommutative
space is the UV/IR mixing\cite{mrs}. In supersymmetric gauge theory in
noncommuting space, linear and quadratic poles in the noncommutativity parameter 
$\tet$ are absent at one loop,
due to the fact that 
contributions from fermionic and bosonic degrees of freedom cancel each other.
First loop Feynman graph calculations for 
noncommutative supersymmetric gauge theory with abelian 
gauge group was studied in \cite{mst}--\cite{ruiz} and $U(N)$ case
was considered in \cite{zanon}--\cite{ferrari2}.

Renormalization of N=1/2 supersymmetric Yang-Mills theory was discussed
in 
\cite{ty},\cite{ghodsi}--\cite{gpr}.
For the gauge group $U(N),$ renormalizability at one loop requires to 
alter the original action.  In \cite{lunin}
it was commented that in supersymmetric gauge theory
where both noncommutativity and
non--anticommutativity are present,
there would be UV/IR mixing.
Although we do not study renormalizability properties of 
the non--anticommutative and noncommutative 
theory, we will show that   UV/IR mixing
is present by 
an explicit calculation for $U(1)$ case.

Seiberg and Witten\cite{sw} introduced an equivalence relation 
between the gauge fields
$\h A$ taking values in noncommutative gauge 
group and the ordinary gauge fields $A$ as
\be
\label{equiv}
\h A (A)+\h\de_{\h\phi}\h A(A) = \h A(A+\de_\phi A) .
\ee
Here $\h\phi$ and $\phi$ denote gauge parameters of the noncommutative and ordinary
cases.  Seiberg--Witten (SW) map allows one 
to deal with noncommutative gauge theory in terms of an action expanded in
the noncommutativity parameter $\tet$ with ordinary gauge fields.
We would like to study its generalization to superspace. 
Gauge transformations of N=1/2 supersymmetric theory in component fields
does not depend on the non--anticommutativity parameter $C,$ owing to the
parametrization of  vector superfield given  in \cite{sei}.
As we will explicitly show,  this is a generalization of SW map to non--anticommutative 
superspace.
Generalizations of SW map to $C$ deformed superspace are studied  in 
\cite{mik} and \cite{ws}. It is also studied in  harmonic superspace \cite{iva},\cite{saf}. We will 
discuss in detail how generalizations of SW map
to superspace can be obtained in terms of  component fields.
When only non--anticommutativity is present one can chose to work either with 
ordinary 
supersymmetry transformations with a deformed gauge transformation or without deforming 
gauge transformations but changing supersymmetry transformations as in \cite{sei}. On the 
other hand for only 
$\theta$ deformed superspace 
SW map includes
deformation of supersymmetry as well as  gauge transformations\cite{pss}--\cite{ku}. 
As  is clear from its definition (\ref{equiv}), SW map refers only to gauge 
transformations.  Hence, it is not a 
priori  guaranteed that a noncommutative gauge theory will be gauge invariant
after performing  SW map. However, as we will show
in the generalization of SW map there is a freedom of choosing
$C\tet$ dependent terms once the $C$ and $\tet$ deformed parts are fixed
separately. Thus, $C\tet$ dependent terms can be chosen 
such that the resultant theory becomes gauge invariant but supersymmetry
transformations should be  deformed. 

In section 2 we present N=1/2 supersymmetric gauge theory action in noncommutative space 
exhibiting its gauge and supersymmetry invariance. Moreover, we show that UV/IR mixing occurs.
In section 3 first we  discuss how to generalize SW map to noncommuting and/or non--anticommuting  
superspace gauge transformations. Then, we apply SW map to $U(1)$ and non--abelian 
 noncommutative and noncommutative gauge theory to obtain $\tet$--expanded
 actions.
Supersymmetry transformations of the latter become non--local to preserve
 gauge invariance of the resultant action.

\section{Noncommutative N=1/2 supersymmetric gauge \mbox{theory}}

In terms of constant, respectively, symmetric and antisymmetric parameters
 $C_{\al\beta}$ and $\trs ,$
let the Grassmann coordinates $\tet_\al ,\ \al=1,2,$ 
$\bar{\tet}_{\dot{\al}} ,\ \dot{\al}=1,2,$
and bosonic coordinates 
$y^\m = \tilde{x}^\m + i\tet \s^\m\bar{\tet},$ $\m =0,\cdots ,3,$
satisfy the deformed brackets\cite{sei}
\bea
\{\h\tet_\al, \h\tet_\bet\} = C_{\al\bet} ,&
{[\h y^\rho ,\h y^\sigma ]}= i\trs ,&\\
\{\h\tet_\al, \bar{\tet}_{\dot{\al}}\} = 0 ,&
\{\bar{\tet}_{\dot{\al}},\bar{\tet}_{\dot{\beta}}\}=0 &\\
{[\h y^\rho ,\bar{\tet}_{\dot{\al}} ]}= 0 ,&
{[\h y^\rho ,\h{\tet}_{{\al}} ]}= 0 .&
\eea
This is possible only in euclidean space.
Although we deal with euclidean
$\mathbb{R}^4,$ we use Minkowski space  notation and follow the
conventions of \cite{wess}.
We will also use the antisymmetric parameter
$C^{\m\n} = C^{\al\bet}\e_{\bet\ga}{\s^{\m\n}_\al}^\ga$ which
satisfies the self--duality property
\be
\label{sdd}
C^{\m\n} = \fr{i}{2} \e^{\m\n\ro\la}C_{\ro\la} .
\ee
An associative star product embracing both of the deformations was introduced
in \cite{ty}
\bea
&f(y,\tet )\ti\star g(y,\tet )  =  f(y,\tet ) 
\exp \Big(
\frac{i}{2} \tet^{\mu\nu} 
\frac{\overleftarrow{\del}}{\del y^\m}
\frac{\overrightarrow{\del}}{\del y^\n} & \no 
&-\frac{1}{2} C^{\al\beta} \frac{\overleftarrow{\del}}{\del \tet^\al}
\frac{\overrightarrow{\del}}{\del \tet^\beta} 
\Big)
g(y,\tet )  
\equiv   f(y,\tet )\ssc \ \sst g(y,\tet ), \label{sct} &
\eea
where the derivatives  $\del /\del \tet^\al$ are defined to be  at fixed $y_\mu$ and
$\bar{\tet}.$ In fact,  one can separate $C$ and $\trs$ dependent parts which we
denote $\ssc $ and $ \sst .$

Instead of dealing with the $\ti\star$ product we will proceed
in a different way.
Seiberg considered the case $\trs =0$ in terms of a
 vector superfield written  in a Wess--Zumino gauge
which was employed to write  the action in commuting 
coordinates $x^\mu$ as (note that we do not deal with $\hat{\tilde{x}}$ coordinates
appearing in $\hat{y}$ which would be noncommuting)
\bea
S &  = &  \fr{1}{g^2} \int {\rm tr} \Big\{ -\fr{1}{4} F^{\m\n}
F_{\m\n} - i \la \sld\lb  + \fr{1}{2} D^2 \no
& & - \fr{i}{2} C^{\m\n}
F_{\m\n} \lb\lb + \fr{|C|^2}{8}(\lb\lb)^2 \Big\} . \label{12}
\eea
$F_{\mu\nu}$ is the
non--abelian field strength related to the gauge field $A_\mu .$
$\la,\ \lb $ are independent fermionic fields and $D$ is
auxiliary bosonic field. Covariant
derivative is defined as $D_\m =\partial_\m +i[A_\m ,\cdot ].$ 
The action (\ref{12}) is invariant under the usual gauge transformations
and it possesses N=1/2 supersymmetry.

Obviously, (\ref{12}) is a theory in commuting coordinates though the
constant parameter $C$ appears. Hence, 
considering it in noncommuting space
letting the coordinates
satisfy 
\be 
[\h x^\m, \h x^\n] = i\tmn
\ee
is legitimate. 
We introduce the star product 
\be
\label{tdef}
f(x) \star g(x) = f(x) e^{\frac{i}{2} \tet^{\mu\nu} \overleftarrow{\del_\m}
\overrightarrow{\del_\n} }g(x) 
\ee
and work with the commuting coordinates $x_\mu $
satisfying the Moyal bracket
\be 
[ x^\m,  x^\n]_\star \equiv x^\mu \star x^\nu -x^\nu \star x^\mu = i\tmn.
\ee

By replacing ordinary products with the star product (\ref{tdef}) in  (\ref{12}),
one obtains  the action
\bea
 I & = &\fr{1}{g^2} \int {\rm tr} \Big\{ -\fr{1}{4}
    \h F^{\m\n} \h F_{\m\n} - i \h\la \sld \star \h{\lb}  + \fr{1}{2} \h D^2 \no
& &- \fr{i}{2} 
C^{\m\n} \h F_{\m\n} \h{\lb}\star\h{\lb} +
    \fr{|C|^2}{8}(\h{\lb}\star\h{\lb})^2 \Big\} .  \label{i2}
\eea
Here we adopted the definitions
\bea
\h{F}_{\m\n} & = & \del_\m \h A_\n - \del_\n \h A_\m + i [ \h A_\m, \h A_\n]_\star ,
\nonumber \\
\sld \star \h{\lb} &=& \ps\h{\lb} + i [ \h{\sla},\h{\lb} ]_\star . \nonumber
\eea

This noncommutative gauge theory would also have resulted from the superfield formulation of N=1/2 supersymmetric theory 
discussed
 in \cite{ty} by making use of the parametrization given in \cite{sei} for vector superfields.
 
We assume that surface terms are vanishing, so that
the following property is satisfied
$$
\int d^4xf(x)\star g(x)=\int d^4xf(x)g(x).
$$

Gauge transformations of the fields are 
\bea
\de \h A_\m &=&  \del_\m \h\phi - i \big[\h\phi, \h
    A_\m\big]_\star ,\nonumber\\
\de \h\la_\al&=& -i\big[\h\phi,\h\la_\al\big]_\star ,\nonumber\\
\de\h{\lb}_{\dot\al} &=& -i\big[\h\phi, \h{\lb}_{\dot\al} \big]_\star ,\nonumber\\
\de\h D &=& -i\big[\h\phi, \h D \big]_\star , \label{sgt}
\eea
where
$\h \phi $ denotes gauge parameter. 
Making use of (\ref{sgt}) one 
can observe the following transformations
\bea
\de\h F_{\m\n} &=& -i\big[\h\phi, \h F_{\m\n} \big]_\star , \nonumber \\
\de (\sld\star\h{\lb})  &=& -i[\h\phi,\sld\star\h{\lb}]_\star\no
\de(\h{\lb}\star\h{\lb})  &=& -i[\h\phi,
\h{\lb}\star\h{\lb}]_\star . \nonumber
 \eea
Therefore, we can conclude that the action (\ref{i2}) is gauge invariant under 
\mbox{noncommutative} $U(N)$ gauge transformations.

On the other hand, 
supersymmetry transformations of the
component fields can be  defined as 
\bea
\de_S\h\la &=& i\xi \h D + \s^{\m\n}\xi
    (\h{F}_{\m\n} + \fr{i}{2}C_{\m\n}\h{\lb}\star\h{\lb}), \label{silk} \\
\de_S \h{A}_\m &=&-i\h{\lb}\bar\s_\m\xi ,\\
\de_S \hat D &=&-\xi\s^\m D_\m\star\h{\lb} , \\
\de_S \h{\lb}&=& 0 ,\label{sson}
\eea
where $\xi^\al$ is a constant Grassmann parameter.
To discuss  supersymmetry properties of the action (\ref{i2}) 
one needs to make use of the relation
$$
\s^{\ro\la}\s^\m = \fr{1}{2} ( - \eta^{\m\la} \s^\ro +
\eta^{\m\ro} \s^\la + i \e^{\m\ro\la\ka} \s_\ka) .
$$
The $C=0$  part can be  shown to be supersymmetric
using the Bianchi identity \mbox{
$\e^{\m\n\la\ro}D_\m \star \h F_{\n\la} = 0,$ } which is 
due to the associativity of star product.
On the other hand the $C_{\m\n}$ dependent terms   yield
\bea
& \int  d^4x \xi\Big\{ 2(\s_\n
        C^{\m\n}D_\m\star\h{\lb})(\h{\lb}\star\h{\lb}) +
        \e^{\m\n\ro\la}\s_\n
C_{\ro\la}(\h{\lb}\star\h{\lb}) & \no
&\times (D_\m\star\h{\lb}) 
- 4 (\s_\n C^{\m\n}D_\m\star\h{\lb})(\h{\lb}\star\h{\lb})
        \Big\} =0,  & \nonumber
\eea
where the self--duality condition (\ref{sdd}) is utilized.
Hence, (\ref{i2}) is a noncommutative  N=1/2 supersymmetric $U(N)$
gauge theory action.

To perform perturbative calculations one should  introduce ghost fields
to fix the gauge. Moreover, matter fields may also be added. 
Let us 
consider noncommutative $U(1)$ gauge group. In this case Feynman rules can be read
from the N=1/2 supersymmetric $U(N)$ gauge theory\cite{ghodsi} by the replacement of the structure constants:
\bea
f_{a_1 a_2a_3} & \longrightarrow & 2\sin \left( \tilde{k}_2 k_3\right) \\
d_{a_1 a_2a_3} & \longrightarrow & 2\cos \left( \tilde{k}_2  k_3 \right),
\eea
where we denoted $\tilde{k}^\mu \equiv \theta^{\mu\nu}k_\nu . $
Here, $k_2$ and $k_3$ are the momenta of the lines corresponding to the indices $a_2$
and $a_3,$ respectively.  Instead of giving a full discussion of one loop calculations,
we would like to consider only the following non--planar one loop diagram

{\begin{eqnarray}
\SetScale{0.8}
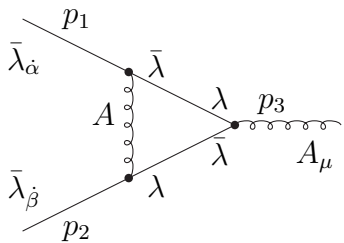
\begin{picture}(-230,45)(0,0)
\Line(-200,50)(-100,0)
\Line(-200,-50)(-100,0)
\Gluon(-150,25)(-150,-25){2}{5}
\Text(-130,4)[]{${A}$}
\Gluon(-100,0)(-50,0){2}{5}
\Vertex(-150,25){2}\Vertex(-150,-25){2}\Vertex(-100,0){2}
\Text(-160,25)[]{${\bar \lambda}_{{\dot \alpha}}$}
\Text(-110,23)[]{${\bar \lambda }$}
\Text(-110,-23)[]{${\lambda }$}
\Text(-86,10)[]{$ \lambda $}
\Text(-86,-10)[]{$ {\bar \lambda} $}
\Text(-160,-25)[]{${\bar \lambda}_{{\dot \beta}}$}
\Text(-50,-12)[]{$A_\mu $}
\Text(-66,8)[]{$p_3$}
\Text(-140,40)[]{$p_1$}
\Text(-140,-40)[]{$p_2$}
\nonumber
\end{picture}
\end{eqnarray} }

\vspace{1cm}

\noindent which is typical of the N=1/2 supersymmetric gauge theory.
The amplitude is proportional to
\bea
& \propto g^3 C^{\kappa\nu}\sigma_{\kappa\beta \dot{\beta}}
\sigma^\mu_{\gamma \dot{\delta}}\epsilon_{\dot{\alpha}\dot{\gamma}}
\int \frac{d^4k}{(2\pi)^4}
\frac{k_\nu (\slk -\slp_1)^{\dot{\gamma}\gamma}
(\slk +\slp_2)^{\dot{\delta}\beta}}{k^2(k-p_1)^2(k+p_2)^2} &\no
& \times  \cos (\tilde{k} p_1) \sin (\tilde{k} p_2) \sin (\tilde{k}  p_3).& \nonumber
\eea
Using the calculation methods of \cite{mst}, one can observe that
this amplitude  produces low momenta poles as
\begin{equation}
g^3 C^{\kappa\nu}\sigma_{\kappa\beta \dot{\beta}}
\sigma^\mu_{\gamma \dot{\delta}}\epsilon_{\dot{\alpha}\dot{\gamma}}
\frac{ \tilde{l}_\nu (\tilde{\sll})^{\dot{\gamma} \gamma }
(\tilde{\sll} )^{\dot{\delta}\beta} } { \tilde{l}^4 },
\end{equation}
where $l_\mu $ are some definite functions of $p:$
$$
l=l(p_1,p_2,p_3).
$$
To get the correct factors we should take into account contributions coming from 
 all of the 
diagrams including ghosts and also matter if they are coupled. Nevertheless,
calculation of the above diagram shows that  UV/IR mixing
occurs.

\section{Generalized SW map and $\mathbf \trs$--expanded action }

To attain $\trs$--expanded action in terms of ordinary component fields we
first should 
discuss  in detail how SW  map  (\ref{equiv}) can be 
generalized to noncommutative and/or non--anticommutative 
superspace. 
SW map (\ref{equiv}) clearly alludes  only to gauge 
transformations, it does not refer to any gauge theory action.
Hence, although one applies the map to a gauge invariant noncommuting and/or 
non--anticommuting theory it is not guaranteed that the resultant action
will possess the ordinary gauge invariance. However,  
we will show that in the superspace 
with noncommuting and 
non--anticommuting coordinates it can be chosen appropriately
such that the resultant action is gauge invariant.

In the ordinary (non--deformed) superspace infinitesimal gauge transformations 
of component fields derived from
\be
\label{inse}
\de_\Lambda e^V=-i \bar\Lambda e^V+ i e^V\Lambda .
\ee
Let us deal with $U(1)$ gauge group to illustrate how
generalizations of SW map can be obtained.
Using the parametrization of   \cite{sei} we define
the non--deformed vector superfield as 
\be
\label{vsff}
V  =  -\tet\sigma^\mu \bar\tet A_\mu + i \tet\tet \bar\tet\lb
-i\bar\tet\bar\tet \tet\la
+\fr{1}{2}\tet\tet\bar\tet\bar\tet \left(D-i\partial_\mu A^\mu\right), 
\ee
which satisfies $ V^2 =  -(1/2)\bar\tet\bar\tet\tet\tet A_\mu  A^\mu $ and $ V^3=0.$ 
The appropriate gauge parameters are 
\begin{eqnarray}
\Lambda & = & \phi + \frac{i}{2}\tet\sigma^\mu \bar\tet \partial_\mu \phi, \\ 
\bar\Lambda & = &\phi - \frac{i}{2}\tet\sigma^\mu \bar\tet \partial_\mu \phi
+\fr{1}{2}\tet\tet\bar\tet\bar\tet \partial^2\phi  .
\end{eqnarray}
To obtain infinitesimal gauge transformations we need to deal not only with
$V$ but with $\Sigma=V+\fr{1}{2}V^2.$ 
Indeed, 
\be
\label{kal}
\de_\Lambda \Sigma =-i \left(\bar\Lambda -  \Lambda
+ \bar\Lambda \Sigma -\Sigma \Lambda \right) ,
\ee
yields the ordinary 
infinitesimal
gauge transformations of the component fields.
Noncommutative gauge transformations are defined as
\be
\de_{\bar\Lambda} \h\Sigma_\Lambda =-i \left(\h{\bar\Lambda} -  \h\Lambda
+ \h{\bar\Lambda} \ti\star \h\Sigma -\h\Sigma \ti\star \h\Lambda \right)  ,
\ee
by replacing 
multiplication of  bilinear components with the star product
$\ti\star ,$ in (\ref{kal}), so that we also need
$$
\h{V} \ti\star \h{V}  =  -\fr{1}{2}\bar\tet\bar\tet\left(\tet\tet\h{A}_\mu \sst\h{A}^\mu
+\fr{1}{4} |C|^2 \h{\lb} \sst \h{\lb} \right) ,
\h{V}\ti\star \h{V} \ti\star \h{V}   =  0.
$$
Note that the star product $\sst$ is in terms of $y$  coordinates.

Generalization of  SW 
map to noncommutative and  non--anticommutative
superspace gauge transformations can be defined  by 
the equivalence relation
\be
\label{equivS}
\h \Sigma (\Sigma )+\h\de_{\h\Lambda }\h \Sigma (\Sigma ) 
= \h \Sigma (\Sigma +\de_\Lambda \Sigma ) .
\ee
This is obtained by replacing the gauge field $A$ with the vector superfield $\Sigma$ and
the gauge parameter $\phi$ with the supergauge parameter $\Lambda$
in (\ref{equiv}). 

We deal with $U(1)$ gauge group to solve the equivalence relation (\ref{equivS}). 
Keeping terms first order in $\tet_{\mu\nu}, C_{\al\beta}$ and $C\theta$
which are denoted as
\bea
\h \Sigma & = & \Sigma +  \Sigma_{(C)}+\Sigma_{(\tet )}
+\Sigma_{(C\tet )}\equiv \Sigma +  \Sigma_{(1)},\\
 \h\Lambda & = & \Lambda +\Lambda_{(C)}+\Lambda_{(\tet )}
+\Lambda_{(C\tet )}\equiv \Lambda +\Lambda_{(1)},\\  
\h{\bar\Lambda} & = & \bar\Lambda + \bar\Lambda_{(C )}
+\bar\Lambda_{(\tet )}+\bar\Lambda_{(C\tet )} \equiv
\bar\Lambda + \bar\Lambda_{(1 )} ,  
\eea
(\ref{equivS}) leads to
\begin{eqnarray}
& \Sigma_{(1)}\left(\Sigma+\del_\Lambda \Sigma \right)  - 
\Sigma_{(1)} \left(\Sigma\right) +i{\bar\Lambda}_{(1)}-i\Lambda_{(1)} 
  = & \nonumber \\
& i \left(\Sigma+\Sigma_{(1)}\right) 
(\ti\star -1) \left(\Lambda + \Lambda_{(1)} \right) & \nonumber \\
& -i\left(\bar\Lambda +{\bar\Lambda}_{(1)}\right)(\ti\star -1)
 \left(\Sigma+\Sigma_{(1)}\right) . &\label{egn}
\end{eqnarray}

To acquire a better understanding let us discuss it
first for only non--anticommutative superspace by
setting $\trs=0 .$
When only $\ssc$ survives (\ref{egn}) leads to
\bea
&\Sigma_{(C)}\left(\Sigma+\del_\Lambda \Sigma \right)   - 
\Sigma_{(C)} \left(\Sigma\right) 
+i{\bar\Lambda}_{(C)}-i\Lambda_{(C)} = & \nonumber \\
& - C^{\al\beta}\left( \del_\al \Sigma \del_\beta \Lambda - 
 \del_\al \bar\Lambda \del_\beta\Sigma\right) , &\label{egnc}
\eea
where we denoted $\del /\del \tet^\al \equiv \del_\al .$
There are two different ways of solving this equation: 
the first one is to choose
\be
\Sigma_{(C)}=0,\ \Lambda_{(C)}=0,\ { \bar\Lambda}_{(C)} = 
\frac{1}{2}\bar\tet\bar\tet \tet_\al C^{\al\beta}\sigma^\mu_{\beta \dot{\al}}\partial_\mu\phi 
\lb^{\dot\al},
\ee
so that the gauge transformations are changed, though
supersymmetry transformations are given by the
ordinary ones. The second
solution is not to retain ordinary gauge transformation but
to  change supersymmetry transformations by deforming the vector superfield as  
\be
\label{ssol}
\Sigma_{(C)}=-\fr{i}{2}\bar\tet\bar\tet \tet_\al C^{\al\beta}\sigma^\mu_{\beta \dot{\al}}A_\mu \lb^{\dot\al},\ \Lambda_{(C)}=0,\ { \bar\Lambda}_{(C)} =0. 
\ee
Indeed, this is the Seiberg's solution which resulted in N=1/2 supersymmetric gauge theory.
In the following we will deal only with the latter solution.

Now, let only $\sst$ survive by setting $C=0$ in (\ref{egn}). Hence,  
in terms of the component fields $V_i\equiv (A,\la ,\lb , D),$ and the ordinary
gauge transformations $\de_\phi ,$ (\ref{egn}) yields
\begin{eqnarray}
A_{(\tet)\mu} (V_i+\de_\phi V_i)-A_{(\tet)\mu} (V_i )& = & 
-\del_\mu \phi_{(\tet)}  \no
&&+ \tet^{\rho \sigma} \partial _\rho \phi \partial_\sigma A_\mu ,
\label{egnti}\\
\la^\al_{(\tet)} (V_i+\de_\phi V_i)-\la^\al_{(\tet)} (V_i ) & =& 
\tet^{\rho \sigma} \partial _\rho \phi \partial_\sigma \la^\al ,\\
\lb^{\dot\al}_{(\tet)} (V_i+\de_\phi V_i)-\lb^{\dot\al}_{(\tet)} (V_i )&  =& 
\tet^{\rho \sigma} \partial _\rho \phi \partial_\sigma \lb^{\dot\al} ,\\
D_{(\tet)} (V_i+\de_\phi V_i)-D_{(\tet )} (V_i ) & = &
\tet^{\rho \sigma} \partial _\rho \phi \partial_\sigma D . \label{egnts}
\end{eqnarray}
They had already been obtained  in \cite{okb}  and can be solved as
\bea  
A_{(\tet) \m} &=&  \tet^{\ro\s}A_\ro( \del_\s A_\m
- \del_\m A_\s /2) ,\label{mmmi} \\
   D_{(\tet)} &=& 
\tet^{\ro\s}A_\ro \del_\s D , \\
\la_{(\tet )\al} &=&
 \tet^{\ro\s} A_\ro \del_\s\la_\al ,\\
\lb_{(\tet)}^{\dot\al} &=&  \tet^{\ro\s}A_\ro 
\del_\s\lb^{\dot\al} .\label{mmmf}
\eea

We are concerned with N=1/2 supersymmetric gauge theory in noncommuting space. Hence,
when we deal with the full--fledged star product $\tilde\star$ we would like to keep 
the Seiberg's solution (\ref{ssol}) for $\Sigma_{(C)}.$
Therefore, we plug (\ref{ssol}) into (\ref{egn}) which results in
\begin{eqnarray}
&\Sigma_{(\tet)}\left(\Sigma+\de_\Lambda \Sigma \right) +
\Sigma_{(C\tet )}\left(\Sigma+\de_\Lambda \Sigma \right) - \Sigma_{(\tet)} \left(\Sigma\right)
&\nonumber \\
&-\Sigma_{(C\tet)} \left(\Sigma\right) 
+i{\bar\Lambda}_{(\tet)}+i{\bar\Lambda}_{(C\tet)}
-i\Lambda_{(\tet)} -i\Lambda_{(C\tet)} 
& \no
& -\fr{1}{2}\bar\tet \bar\tet \Big[ \tet\tet (
\partial_\mu\phi_{(\tet)}A^\mu  
+ \partial_\mu\phi A^\mu_{(\tet)} 
+\partial_\mu\phi_{(C\tet)}A^\mu 
& \no
&+ \partial_\mu\phi A^\mu_{(C\tet)} ) 
 +i \tet_\al C^{\al\bet}\sigma_{\beta \dot{\al}}^\mu
(\partial_\mu \phi_{(\tet)} \lb^{\dot{\al}} 
+ \partial_\mu \phi \lb^{\dot{\al}}_{(\tet)} )
\Big] & \no
& = \fr{i}{4} \trs C^{\al\beta}
\left( \del_\al\partial_\rho\bar\Lambda  
\del_\beta \partial_\sigma \Sigma
-\del_\al \partial_\rho\Sigma
\del_\beta \partial_\sigma \Lambda
\right) .& \label{cnt}
\end{eqnarray}
For only $\trs$ part we  would like to retain the equations (\ref{egnti})--(\ref{egnts})
by adopting the solutions (\ref{mmmi})--(\ref{mmmf}).
After some calculations, one can show that (\ref{cnt}) simplifies and 
we are left with only $C\tet$ dependent terms:
\bea
&\Sigma_{(C\tet )}\left(\Sigma+\de_\Lambda \Sigma \right) 
-\Sigma_{(C\tet)} \left(\Sigma\right) 
+i{\bar\Lambda}_{(C\tet)}-i\Lambda_{(C\tet)} &\no
&-\fr{1}{2}\bar\tet \bar\tet \tet\tet (
\partial_\mu\phi_{(C\tet)}A^\mu 
+ \partial_\mu\phi A^\mu_{(C\tet)} 
)  =0.& \label{nlc}
\eea
In components it yields
\begin{eqnarray}
A^\m_{(C\tet )} (V_i+\de_\phi V_i)-A^\m_{(C \tet)} (V_i )+\del^\mu 
\phi_{(C\tet)} & = & 0,
\label{mi} \\
\la^{\al}_{(C\tet )} (V_i+\de_\phi V_i)-\la^{\al}_{(C\tet )} (V_i ) & = &0,\\
\lb^{\dot\al}_{(C)} (V_i+\de_\phi V_i)-\lb^{\dot\al}_{(C\tet )} (V_i ) & = & 0,\\
D_{(C\tet )} (V_i+\de_\phi V_i)-D_{(C\tet )} (V_i ) & = & 0.\label{ml}
\end{eqnarray}
We will not work out their solutions\footnote{In 
\cite{mik} after making use of Seiberg`s solution 
for $C$ deformed vector superfield (\ref{ssol}), additional
 $C$ dependent fields and parameters were  introduced
 and equations for components were derived. They are the same with
(\ref{mi})--(\ref{ml}) by
replacing $C\tet$ components of fields and 
parameters with these additional $C$ dependent ones. 
Hence, the discussions 
of \cite{mik} regarding solutions can be applied to our case 
up to an overall $\trs$ dependence. }
but note that there is the trivial solution 
\be
\label{triv}
A^\m_{(C\tet)} =\la^\al_{(C\tet)}=\lb^{\dot\al}_{(C\tet)}
=D_{(C\tet)}=\phi_{(C\tet)}=0.
\ee
Instead of dealing
with the trivial solution (\ref{triv}) one can add  $C\tet$ dependent terms. 
However, this will result in changing the supersymmetry transformations of
the fields as will be discussed.

When component fields are deformed by adding $C$ and/or $\trs$ dependent terms,
supersymmetry transformations of commuting fields should also be 
deformed,\cite{sei},\cite{pss}--\cite{ku}. 
Let us discuss how deformed supersymmetry transformations can be obtained.
Denote the original 
supersymmetry transformations of components  as
\be
\label{orsut}
\dss V_i=f_i(V_j,\xi ).
\ee
By replacing the original  components with the deformed ones  one gets
\be
\label{orsu}
\dss \h{V}_i=f_i(\h{V}_j,\xi ).
\ee
Now, perform the map
\be
\h{V}_i (V) =V_i+V_{i(C)}+V_{i(\tet)}+V_{i(C\tet)},
\ee
and plug  it in the left as well as in the right hand side of (\ref{orsu}).
Then
one can read deformed supersymmetry transformation of ordinary component fields as
\be
\label{deff}
\dss V_i=f_i(\h V_j(V),\xi )-\dss V_{i(C)}-\dss V_{i(\tet)}- \dss V_{i(C\tet)}.
\ee
As an example one can show that 
when one deals with only $C$ deformed case and adopts
the Seiberg's solution  (\ref{ssol}), the N=1/2 supersymmetry transformations
(\ref{silk})--(\ref{sson}) with $\trs =0$ follow. 

Let us apply SW map to N=1/2 supersymmetric gauge theory in noncommutative 
space. For $U(1)$ case 
we adopt the Seiberg's solution for the $C$ dependent part (\ref{ssol}),
hence the action (\ref{i2}) with $U(1)$ results. Then,
using the solutions (\ref{egnti})--(\ref{egnts})
and (\ref{triv})  for $\trs$ and $C\tet$ dependent terms, (\ref{i2})
yields the $\tet$--expanded N=1/2 supersymmetric
gauge theory action up to the
first order in $\tet $ 
for $U(1)$  
\bea 
&I^{(1)} = \int d^4x \Bigg[
-\fr{1}{4}F_{\m\n}F^{\m\n}  - i \la\ps\lb -
\fr{i}{2}C^{\m\n}F_{\m\n}{\lb}^2 & \no
&+ \fr{1}{2}D^2- 
\tet^{\ro\s}\Bigg( -\fr{1}{2}F^{\m\n}F_{\n\s}F_{\m\ro} +
\fr{1}{8}F_{\ro\s}F_{\m\n}F^{\m\n} 
& \no
&- \fr{1}{4}D^2F_{\ro\s}
 +\fr{i}{2}F_{\ro\s}\la\ps\lb  + i\la\s^\m\del_\s\lb F_{\m\ro} &\no
&  +\fr{i}{2}C^{\m\n}F_{\m\ro}F_{\n\s}\lb^2 -
\fr{i}{4}C^{\m\n}F_{\ro\s}F_{\m\n}{\lb}^2 \Bigg)  \Bigg] .&\label{te1}
\eea
Although (\ref{te1}) possesses the usual $U(1)$ gauge invariance its
supersymmetry transformations should be altered.
Deformed supersymmetry transformations can be read by making use of the
general formula (\ref{deff}) with
the transformations (\ref{silk})--(\ref{sson}) as
\bea 
     \de_S A_\m &=& i\xi\s_\m\lb -\fr{i}{2}\trs\xi\s_\ro\lb(\del_\s A_\m + F_{\s\m}) +\no
&&\fr{i}{2}\trs\xi\s_\s A_\ro\del_\m\lb ,\label{asus}\\
     \de_S \la &=& i\xi D - \xi\s^{\m\n}F_{\m\n} + \trs\xi\s^{\m\n}F_{\m\ro}F_{\n\s} \no
&&+\fr{i}{2}\s^{\m\n}\xi C_{\m\n}\lb^2 - i \xi \trs\del_\ro\la\s_{\s}\lb , \\
\de_S\lb &=&
-i\trs\xi \del_\ro\lb\s_{\s}\lb ,\\
\de_S D &=& -\xi\s^\m\del_\m\lb + \trs\xi\s^\m\del_\ro\lb F_{\m\s} \no
&&+i\trs\xi\s_\s\del_\ro D\lb , \label{dsus}
\eea
which can be shown to yield
\bea
\de_S F_{\m\n} & =& i\xi(\s_\n\del_\m\lb - \s_\m\del_\n\lb) +
i\xi \trs \s_\ro (\del_\m\lb F_{\n\s} \no
&& - \del_\n\lb F_{\m\s})- i\xi\trs\s_\ro\lb \del_\s F_{\m\n} . \nonumber
\eea
In fact, 
we explicitly checked that the action (\ref{te1})  is invariant under the
$\tet$--expanded supersymmetry transformations  (\ref{asus})--(\ref{dsus}).

The $\tet$--expanded $U(1)$ gauge theory action (\ref{te1}) can be utilized to study  
some different aspects of noncommuting N=1/2 supersymmetric gauge theory. 
Similar to noncommuting electrodynamics one can calculate one loop renormalization properties of this theory\cite{wul} and find solutions of equations of motion\cite{nces}. Moreover, using the master action of
N=1/2 supersymmetric $U(1)$ gauge theory given in \cite{olk} one can study duality properties of the action (\ref{dsus}).

We would like to apply SW map to non--abelian gauge theory 
in the light of the approach used for $U(1).$ 
Thus we adopt Seiberg's solution for only $C$ dependent 
part of SW map which yields the noncommutative $U(N)$ gauge theory action (\ref{i2}). Then,
for the $\trs$ part of the component fields, 
we adopt the  generalization of SW map given  in 
\cite{w1}:
\bea A_{(\tet)\m} &=&  \fr{\tet^{\ro\s}}{4}\{A_\ro, \del_\s A_\m
+ F_{\s\m}\} ,\no 
F_{(\tet)\m\n} &=& 
\fr{\tet^{\ro\s}}{4}\left( 2\{F_{\m\ro}, F_{\n\s}\} - \{A_\ro,
(D_\s+\del_\s)F_{\m\n}\} \right) ,\no  
D_{(\tet)} &=& 
\fr{\tet^{\ro\s}}{4}\{A_\ro, (D_\s + \del_\s)D\} ,\no 
\la_{(\tet)\al} &=&
 \fr{\tet^{\ro\s}}{4}\{A_\ro, (D_\s+\del_\s)\la_\al\} ,\no
\lb_{(\tet)}^{\dot\al} &=&  \fr{\tet^{\ro\s}}{4}\{A_\ro,
(D_\s + \del_\s)\lb^{\dot\al}\} .\label{mmm}
\eea
As we have  already emphasized SW map does not refer 
to any action but it is an equivalence relation between gauge transformations. Hence, a 
priori one cannot guarantee that a noncommutative gauge theory will remain gauge invariant 
under SW map. Indeed, if we choose to work with the trivial solution (\ref{triv}) for $C\tet$
dependent terms, resultant theory will not be gauge invariant. Therefore,
as the  $C \tet $ term we  choose the 
 non--local one 
\be
\label{nlt}
\la_{(C\tet)}^{\al} =
-\fr{\trs}{8}C^{\mu\nu}F_{\m\n}
\{[\lb_{\dot\al},A_\ro],(\partial_\s + D_\s ) 
\lb^{\dot\al}\} (\sigma^\kappa D_\kappa \lb)^{-1}_\al ,
\ee
with the other components vanishing. Obviously, the gauge
parameters $\Lambda_{(C\tet)}$ and $\bar{\Lambda}_{(C\tet)}$ 
should be appropriately chosen such that (\ref{nlc}) be  satisfied. 
When we employ the map (\ref{mmm}) and (\ref{nlt})
in the action (\ref{i2}), up to some surface terms, we attain 
\bea
&I = {\rm tr} \Bigg[ -\fr{1}{4} F^{\m\n}F_{\m\n} 
    - i\la\s^\m D_\m\lb + \fr{1}{2}D^2 &\no 
&+\fr{\trs}{8}\Big( 4F^{\m\n}F_{\m\ro}F_{\n\s} 
-F_{\s\ro}F^{\m\n}F_{\m\n} +2D^2 F_{\ro\s} &\no
&- 2\{F_{\ro\s}, \la\}\s^\m D_\m\lb 
-4\la\s^\m \{F_{\m\ro}, D_\s\lb \}\Big) & \no
&- \fr{i}{2}C^{\m\n}\Big( F_{\m\n}\lb^2 - \trs
F_{\m\ro}F_{\n\s}\lb^2
- \fr{\trs}{4}\{F_{\s\ro}, F_{\m\n}\}\lb^2 & \no
    & + 
\fr{|C|^2}{8}\Big(\lb^2\lb^2 - \fr{\trs}{4} \{F_{\s\ro}, \lb^4\}\Big)
\Bigg] .&
\eea

The price which we pay for 
adding some $C\tet$ dependent terms to obtain gauge invariance is to change supersymmetry
transformations in terms of (\ref{deff}).
Obviously, new supersymmetry transformations 
of $\la$ will have a non--local part. 
For non--abelian case even the local parts of new supersymmetry transformations   
become so
complicated that we do not present them here. Obtaining some other solutions
of (\ref{nlc}) which respects gauge invariance is an open problem which should be
studied.

\vspace{1cm}

\begin{center}
{\bf Acknowledgments}
\end{center}

We would like to thank K. \"{U}lker for fruitful discussions.

\end{document}